\documentclass[prl,twocolumn,floatfix,preprintnumbers,amsmath,amssymb,superscriptaddress]{revtex4}

\usepackage{graphicx}
\usepackage{dcolumn}
\usepackage{amsfonts}
\usepackage{bm}

\begin{document}

\title{Magnetic and orbital ordering in the spinel MnV$_{2}$O$_{4}$}

\author{V. O. Garlea}
 \email{garleao@ornl.gov}
\affiliation{Neutron Scattering Science Division, Oak Ridge National Laboratory, Oak Ridge, Tennessee 37831, USA.}
\author{R. Jin}
\affiliation{Materials Science and Technology Division, Oak Ridge National Laboratory, Oak Ridge, Tennessee 37831, USA.}
\author{D. Mandrus}
\affiliation{Materials Science and Technology Division, Oak Ridge National Laboratory, Oak Ridge, Tennessee 37831, USA.}
\author{B. Roessli}
\affiliation{Laboratory for Neutron Scattering ETHZ \& Paul Scherrer Institute, CH-5232 Villigen PSI, Switzerland}
\author{Q. Huang}
\affiliation{NIST Center for Neutron Research, Gaithersburg, Maryland 20899, USA}
\author{M. Miller}
\affiliation{IPNS Division, Argonne National Laboratory, Argonne, Illinois 60439, USA.}
\author{A. J. Schultz}
\affiliation{IPNS Division, Argonne National Laboratory, Argonne, Illinois 60439, USA.}
\author{S. E. Nagler}
\affiliation{Neutron Scattering Science Division, Oak Ridge National Laboratory, Oak Ridge, Tennessee 37831, USA.}

\date{\today}

\begin{abstract}
Neutron inelastic scattering and diffraction techniques have been used to study the MnV$_{2}$O$_{4}$ spinel system. Our measurements show the existence of two transitions to long-range ordered ferrimagnetic states; the first collinear and the second noncollinear. The lower temperature transition, characterized by development of antiferromagnetic components in the basal plane, is accompanied by a tetragonal distortion and the appearance of a gap in the magnetic excitation spectrum. The low-temperature noncollinear magnetic structure has been definitively resolved. Taken together, the crystal and magnetic structures indicate a staggered ordering of the V $d$ orbitals. The anisotropy gap is a consequence of unquenched V orbital angular momentum.
\end{abstract}


\maketitle

Understanding the consequences of orbital degeneracy, and the interplay of spin, orbital and lattice degrees of freedom, has emerged as a forefront area of condensed matter physics.  One of the most investigated prototypical systems in which these effects are important is the vanadium oxide spinel, with formula $A$V$_{2}$O$_{4}$. There has been much experimental and theoretical effort to understand the properties of $A$V$_{2}$O$_{4}$~\cite{mgv2o4,znv2o4,cdv2o4,Tsunetsugu,Motome,Motome-jpsj,Tchernyshyov,DiMateo,Maitra}, where $A$ is a non-magnetic species such as Mg~\cite{mgv2o4}, Zn~\cite{znv2o4}, or Cd~\cite{cdv2o4}. As is well known, the V$^{3+}$ ($3d^2$) ion sits in a position of local octahedral symmetry and therefore has a threefold degenerate orbital ground state. Furthermore, the V$^{3+}$ ions occupy the vertices of a tetrahedron and their mutual antiferromagnetic superexchange interactions are topologically frustrated. A common feature found in these materials is a sequence of two phase transitions~\cite{mgv2o4,znv2o4,cdv2o4}.  The higher temperature transition is a structural distortion involving a compression of the VO$_{6}$ octahedra and a consequent partial lifting of the orbital degeneracy. The orbital ordering is accompanied, at lower temperature, by an antiferromagnetic ordering.  Various models~\cite{Tsunetsugu,Motome,Motome-jpsj,Tchernyshyov,DiMateo,Maitra} have been proposed to explain this behavior. However, up to date, there is not yet a full consensus on the precise nature of the orbital ordering.

Replacing the atom at the $A$ site by a magnetic species changes the physics, leading to different and very interesting behavior.  Recent attention has turned to MnV$_{2}$O$_{4}$~\cite{Plumier,Adachi,Suzuki,Baek}, where the $A$ site ion, Mn$^{2+}$, is in a $3d^5$ high spin configuration $S$=5/2 with quenched orbital angular momentum.  MnV$_{2}$O$_{4}$ exhibits a phase transition at approximately 56~K from a paramagnetic cubic phase into a collinear ferrimagnetic phase, also with a cubic structure. Around 53~K there is a second transition to a tetragonal structure, with the spin structure becoming non-collinear~\cite{Plumier}. Recently, it was found that the cubic to tetragonal transition could be induced by a modest magnetic field of a few Tesla~\cite{Adachi}. Since then there have been careful x-ray scattering studies of the structure showing that the low temperature space group is $I4_{1}/a$~\cite{Suzuki}. In addition, it has been suggested based on NMR and susceptibility measurements that the system also exhibits a re-entrant spin glass behavior~\cite{Baek}. In this paper we report the results of neutron scattering measurements on both powder and single crystal samples of MnV$_{2}$O$_{4}$. Neutron measurements show the existence of two phase transitions, and allow for a definitive determination of the low temperature non-collinear ferrimagnetic structure. Inelastic neutron scattering shows that the cubic to tetragonal transition is associated with the opening of a gap in the magnetic excitation spectrum. These observations put tight constraints on theoretical models for the spin and orbital physics of MnV$_{2}$O$_{4}$.

\begin{figure}[btp]
\includegraphics[width=3.5in]{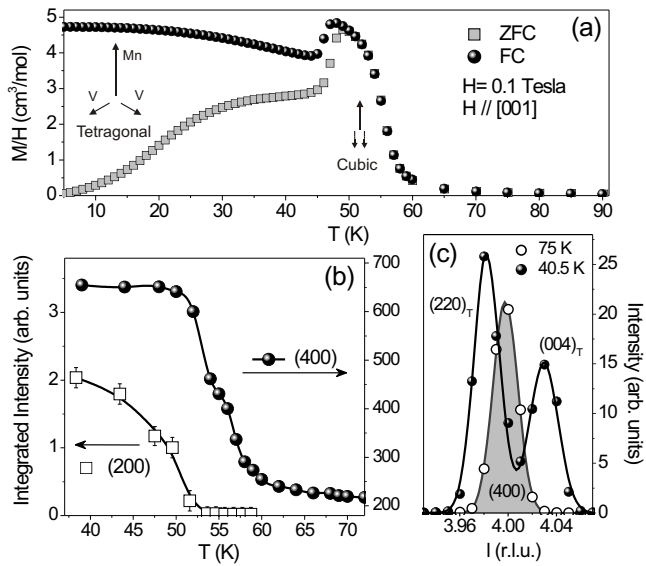}
\caption{\label{orderparam}(a) The zero-field-cooling (ZFC) and the field-cooling (FC) temperature dependence of magnetization in MnV$_{2}$O$_{4}$ single crystal. (b) Temperature dependence of the (400) and (200) Bragg peaks integrated intensities showing the existence of two magnetic transitions. The solid lines are guides to the eye. (c) Splitting of the (400) cubic peak into two tetragonal (220)$_T$ and (004)$_T$, as the crystal structure distorts to form a body-centered tetragonal phase.}
\end{figure}

The powder sample used in this study was prepared by solid-state reaction from stoichiometric mixture of MnO and V$_2$O$_3$. MnV$_{2}$O$_{4}$ single crystals were grown using the floating-zone technique. All samples were characterized by x-ray diffraction and magnetization measurements. The temperature dependence of the low field (0.1~kOe) magnetization of a MnV$_{2}$O$_{4}$ single crystal under zero-field cooling (ZFC) and field cooling (FC) conditions is shown in Fig.~\ref{orderparam}(a). Qualitatively, the magnetization curves look similar to those reported previously~\cite{Suzuki,Baek}. Upon decreasing temperature, the FC magnetization  goes through a maximum and shows a sharp decrease before it starts to increase again. As visible in Fig.~\ref{orderparam}(a), the rapid decrease in magnetization, near 50~K, marks the point where the FC and ZFC curves begin to diverge. Below 50~K, the ZFC magnetization shows a gradual decrease, while the FC magnetization increases.

To monitor the changes in the crystal and magnetic structures across these transitions, several neutron scattering experiments were performed. Time-of-flight neutron diffraction measurements were carried out using the single crystal diffractometer (SCD) at the IPNS, Argonne, on a 0.2~g single-crystal specimen. High-resolution neutron powder diffraction measurements were performed on the BT1 diffractometer at NCNR, using the wavelength 1.54 {\AA}. Additional elastic and inelastic measurements were conducted using the cold-neutrons 3-axis spectrometer TASP, at the SINQ spallation source. For these, we made use of a single-crystal of approximately 1.3~g, aligned in the $(hhl)$ horizontal scattering plane. Elastic measurements were carried out using a monochromatic neutron beam (with wave-vector $k$ = 1.97~\AA$^{-1}$) and $20'$ collimators in front and after the sample. For the inelastic neutron scattering (INS) measurements, TASP spectrometer was operated with a fixed final energy $E_f$ = 3.5~meV, open$-40'-40'-80'$ collimation, and a cooled Be filter after the sample.

\begin{figure}[tp]
\includegraphics[width=3.5in]{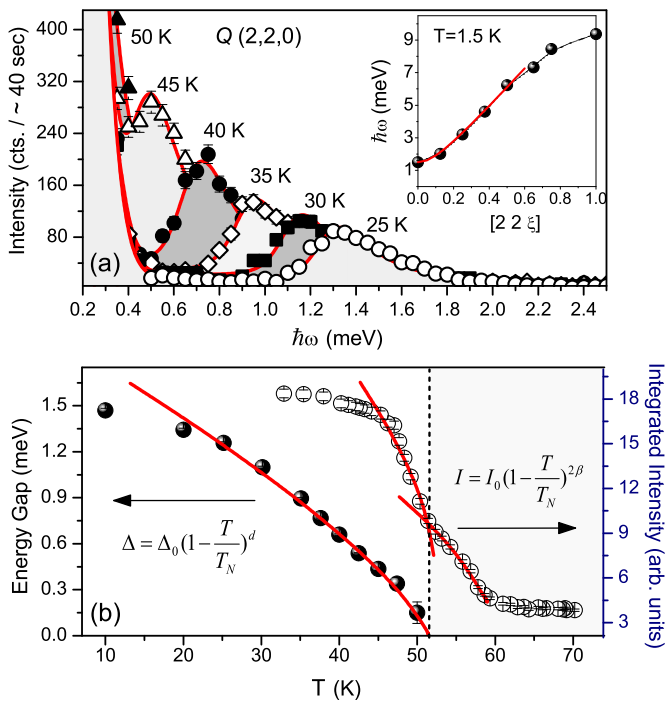}
\caption{\label{gap}(a) Energy-scans measured at different temperatures, at Q = (2 2 0). Solid lines are fits to the data. The large peak centered on 0 meV is due to incoherent scattering. Inset: Dispersion of the low-energy spin-wave branch along the $c$-axis, measured at 1.5~K. (b) Variation with the temperature of the energy-gap (filled dots) and (220) peak intensity (open dots). The gap mode extrapolates to zero at the same temperature where the structural transition occurs. Power-law fit, shown by solid lines, indicate that $\Delta(T)$ scales approximately as the square of staggered magnetization.}
\end{figure}

Fig.~\ref{orderparam}(b) shows the evolution with temperature of the (400) and (200) peaks integrated intensities, measured using the SCD. Diffraction peaks are indexed in the cubic-spinel structure. The (400) peak intensity starts to increase at approximately 60~K ($T_F$) and exhibits a kink at about 52~K ($T_S$). As the temperature is lowered, the intensity continues to increase and saturates below 45~K. It is noteworthy that the order-parameter profile shows a long tail which extends above 70~K. This may be due to the critical scattering associated with a short-range order above $T_F$. A very similar temperature dependence was observed for the (220) peak, measured using the 3-axis spectrometer (Fig.~\ref{gap}(b)). In contrast, the (200) reflection, which is forbidden in the cubic symmetry ($Fd\overline{3}m$), appears only below $T_S$ and increases as the temperature decreases. Further examination of diffraction data confirmed that the transition at $T_S$ is related to the modification of the crystal structure. The occurrence of tetragonal distortion is clearly demonstrated in the Fig.~\ref{orderparam}(c) by the splitting of the (400) Bragg peak into two components, which in the tetragonal unit cell ($a_{T} \approx a/\sqrt{2}$ and $c_{T} \approx c$) can be indexed as (220)$_T$ and (004)$_T$ .

\begin{figure}[tp]
\includegraphics[width=3.45in]{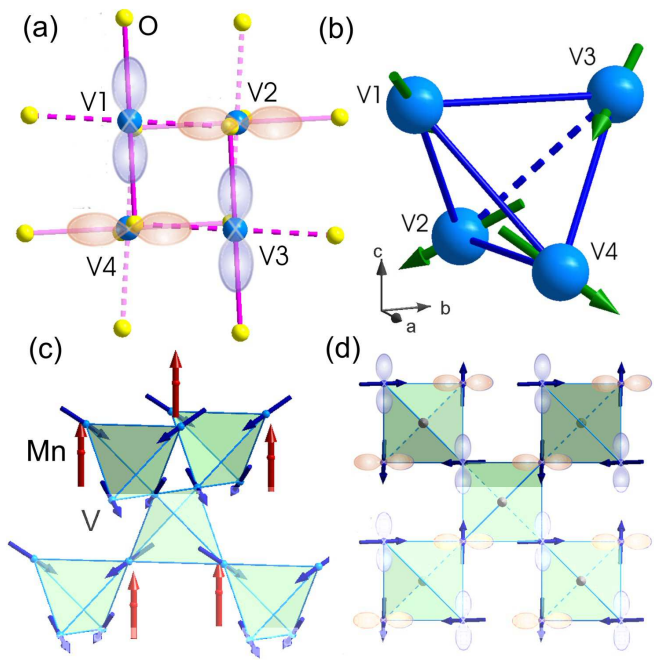}
\caption{\label{model}(Color online) (a) Projection of the V$_4$ tetrahedron in the $ab$ plane. V-O bonds are arranged in an alternating pattern giving rise to a staggered-like orbital arrangement. (b)(c) Graphical representations of the low-temperature non-collinear ferrimagnetic structure of the MnV$_{2}$O$_{4}$. The Mn moments are aligned parallel to the $c$ axis, while the V moments are canted by approximately 65$^{\circ}$. (d) Projection of the magnetic structure on the tetragonal basal plane.}
\end{figure}

The low-temperature crystal structure was determined from a complete set of SCD data that mapped the entire reciprocal space. Mixed nuclear and magnetic reflections with $Q$ ($=4\pi\sin\theta/\lambda$)$\leq$ 7~{\AA}$^{-1}$ were excluded from the structure refinement, carried out using GSAS~\cite{VonDreele}. Checking the systematic absences among high-$Q$ reflections, we confirmed the $I4_{1}/a$ space group, recently proposed by Suzuki \textit{et al.}~\cite{Suzuki}. This space group implies a relaxation of the symmetry constraints at the oxygen site (Wyckoff positions: $16f$). Thus, the compression of the VO$_6$ octahedra along the $c$ crystallographic axis ($c_{T}/a_{T}\approx0.98$) is accompanied by a small distortion of the V-O bonds in the basal plane. Referring to Fig.~\ref{model}(a) which shows a view of the V-tetrahedron along the $c$ axis, the V-O bonds are arranged in an antiferrodistortive alternating pattern with short (dashed lines) and long (bold lines) distances. This is consistent with the so-called $A$-type orbital ordering with antiferro-order along the $c$ axis and ferro- in the $ab$ plane, similar to that proposed by Tsunetsugu and Motome~\cite{Tsunetsugu,Motome} for V-spinels with nonmagnetic $A$-site cations.

Between the two transition temperatures ($T_F$,$T_S$), the magnetic scattering appears only on top of the structural reflections, validating the collinear ferrimagnetic model proposed by Plumier and Sougi~\cite{Plumier}. On the other hand, one observes below $T_S$ the appearance of additional magnetic peaks: ($hk0$) with $h,k\neq2n$, symmetry-forbidden in the tetragonal space-group $I4_{1}/a$. For instance, the cubic (200) peak, indexed in the tetragonal lattice as (110)$_T$, is purely magnetic and can be explained by an antiferromagnetic (AFM) ordering within the $ab$ plane. Nevertheless, the $c$-axis components of Mn and V moments remain ferrimagnetically ordered. To generate all the spin configurations compatible with the crystal symmetry we carried out a group theory analysis~\cite{Kovalev} using the programs SARA$h$~\cite{Sarah} and BasIREPS~\cite{Software}. There are eight irreducible representations (IR) associated with the $I4_{1}/a$ space group and $\roarrow {k}$=(000). Among these, only one allows for a ferrimagnetic alignment of Mn and V sublattices. The basis vectors of this representation are listed in Table~\ref{Table}. It shows that the Mn moments are aligned parallel to the tetragonal $c$-axis, while the V moments may have components on any of the three crystal axes. Interestingly, the model proposed by Plumier~\cite{Plumier} with V-moments lying in ($h$00) sheets, is excluded by symmetry. The $a$ and $b$-axis components are constrained to form an orthogonally stacked AFM structure as shown in Fig.~\ref{model}.

Full refinements of various symmetry-allowed magnetic structure models were performed using high-resolution powder diffraction data.  The use of powder sample avoids complications related to multiple magnetic domains and extinctions. Rietveld refinements were performed using the FULLPROF program~\cite{Fullprof}. In the powder diffraction investigation, the cubic phase was found to persist in a small fraction below $T_S$, and it has been taken into account in the refinements.

\begin{table}[tp]
\caption{\label{Table}Basis vectors (BVs) of an IR of the tetrahedral space group
$I4_{1}/a$ (with $\roarrow k$ = (0,0,0)). BVs are defined relative to the tetragonal axes. Magnetic moments for an atom $j$ is given by m$_j$=$\underset{i}\sum{C_i\psi_i}$ where $C_i$ is the mixing coefficient of BVs $\psi_i$.}
\begin{ruledtabular}
\begin{tabular}{cccccccccc}
& \multicolumn{3}{c}{$\psi_1$}&\multicolumn{3}{c}{}&\multicolumn{3}{c}{} \\[3pt]
Mn1~(0,$\frac{1}{4}$,$\frac{1}{8}$) &(0 &0 &1)&&&\\[3pt]
Mn2~($\frac{1}{2}$,$\frac{1}{4}$,$\frac{3}{8}$) &(0 &0 &1)&&&\\[3pt]
\hline& \multicolumn{3}{c}{$\psi_1'$}&\multicolumn{3}{c}{$\psi_2'$}&\multicolumn{3}{c}{$\psi_3'$} \\[3pt]
V1~(0,0,$\frac{1}{2}$) &(0 &0 &-1)&(1 &0 &0)&(0 &1 &0)\\[3pt]
V2~($\frac{1}{4}$,$\frac{3}{4}$,$\frac{1}{4}$)&(0 &0 &-1)&(0 &1 &0)&(-1 &0 &0)\\[3pt]
V3~(0,$\frac{1}{2}$,$\frac{1}{2}$)&(0 &0 &-1)&(-1 &0 &0)&(0 &-1 &0)\\[3pt]
V4~($\frac{3}{4}$,$\frac{3}{4}$,$\frac{1}{4}$) &(0 &0 &-1)&(0 &-1 &0)&(1 &0 &0)\\[3pt]
\end{tabular}
\end{ruledtabular}
\end{table}

Above the structural transition, the diffraction pattern was fitted using a simple collinear ferrimagnetic model, with the Mn and V moments aligned antiparallel to each other. Such a collinear order may be stabilized by an order-by-disorder mechanism~\cite{Villain}, as previously suggested in Ref.~\cite{Motome}. As discussed above, below $T_S$ the V-moments develop AFM components parallel to the $ab$ plane. As the Mn moments produce a strong effective magnetic field for the V ions, the V moments tend to orient themselves almost perpendicular to the direction of the effective field. The magnetic structure model presented in Table~\ref{Table} gave the best fit to the diffraction pattern. Refinements using spherical Mn$^{2+}$ and V$^{3+}$ form factors, on the 5~K data, yielded the ordered moments: $m_{Mn}\approx4.2~\mu _{B}$ and $m _{V}\approx1.3~\mu _{B}$. The V moments are canted with respect to the $c$-axis by approximately 65.12$^{\circ}$. A stereographic view of the magnetic structure is illustrated in Fig.~\ref{model}(b)(c). The $ab$-projections of V moments are antiparallel within each layer and orthogonal between layers. According to Kugel-Khomskii's prediction~\cite{Kugel}, this arrangement may favor the $A$-type orbital order. Fig.~\ref{model}(d) displays a projection of the MnV$_{2}$O$_{4}$ structure in the $ab$ plane and the compatibility between the magnetic and a staggered orbital ordering.

The structural investigation was complemented with inelastic neutron scattering measurements performed at different temperatures ranging from 1.5~K to 70~K. The spin-wave spectrum across the Brillouin zone was measured in a series of constant-$Q$ scans. At 1.5 K, the spectrum consists of a gapped acoustic mode (with the gap $\Delta\approx1.5~meV$) and several optical branches. The dispersion of the low-energy mode along the $c$-axis is shown in the inset of Fig.~\ref{gap}(a). A detailed analysis of the spectrum will be reported elsewhere~\cite{Garlea}. Here we will focuss only on the temperature behavior of the energy gap. Fig.~\ref{gap}(a) shows the energy scans measured at the zone center (220), at various temperatures. Fits of the gap mode, indicated by solid lines in Fig.~\ref{gap}(a), were done using a cross section function:
\begin{equation}
 \frac{d^2 \sigma}{d \Omega d E'} \propto \frac{\mathbf{A_q}}{\omega_\mathbf{q}} \mathbf{L}(\omega-\omega_\mathbf{q},\Gamma)[n(\omega_\mathbf{q})+1] \label{sqw}
\end{equation}
where $\mathbf{A_q}$ term describes the q-dependence of the intensity, \textbf{L} is a Lorentzian peak-shape function of width $\Gamma$ and $[n(\omega_\mathbf{q})+1]$ is the Bose factor. The approximate dispersion relation near the zone center was written as: $\omega_\mathbf{q}^2 = \Delta^2 + (v\mathbf{q})^2$, where $v$ is the spin wave velocity determined to be approximately 1.88 meV. The model cross section was numerically convoluted with the spectrometer resolution function calculated in the Popovici approximation~\cite{Popovici}.

The variation of the energy gap, $\Delta$, with temperature is shown in Fig.~\ref{gap}(b). It decreases with increasing temperature and vanishes, moving into the incoherent peak, exactly at the structural transition ($T_S$). The energy gap in the spin-wave spectrum can be explained by the existence of single-ion anisotropy. Such anisotropy may be a consequence of the unquenched orbital angular momentum of the V$^{3+}$ ion. To quantify the temperature dependence of the gap, we performed a least-square fit  using the power-law $\Delta(T)\propto(T_S-T)^{d}$. The fit yielded an exponent $d\simeq$ 0.73(2). For comparison, Fig.~\ref{gap}(b) displays the temperature dependence of the (220) Bragg peak intensity, measured simultaneously with the inelastic scans. Near the two transitions, the variation with temperature of the order parameter was described as $I(T)\propto (T_{F,S}-T)^{2\beta}$. The critical exponents reproducing the temperature dependence of the (220) peak are $\beta_1\simeq$ 0.32(4) at the first transition, and $\beta_2\simeq$ 0.34(5) for the second. Although the fits were done over a limited number of data points, one should point out that the obtained values are close to those of the 3D Heisenberg ($\beta=0.36$) or 3D Ising ($\beta=0.33$) models. Comparing these critical exponents with that measured for $\Delta(T)$, shows that the energy of the gap mode varies roughly as the square of staggered magnetization.

In summary, our neutron scattering measurements on MnV$_{2}$O$_{4}$ show the existence of two consecutive magnetic transitions. At about 60~K the transition is from a paramagnetic to collinear ferrimagnetic state. At 52~K the ferrimagnetic state becomes noncollinear, with V spins developing AFM components in the $ab$ plane. There is a simultaneously structural distortion to an orbitally ordered tetragonal phase. The appearance of an anisotropy gap indicates a strong influence of the V orbital angular momentum. A proper theory for spin and orbital physics of MnV$_{2}$O$_{4}$ should account for these observations.

The authors thank D. Abernathy, O. Tchernyshyov and M. Whangbo for valuable discussions and interest in this work. Work at ORNL was supported by the U.S. Department of Energy (DOE) under Contract No. DE-AC05-00OR22725 with UT-Battelle, LLC. Work at Argonne National Laboratory was supported by the US DOE, Office of Science, Office of Basic Energy Sciences, under contract DE-AC02-06CH11357. We also acknowledge the support of the NIST, U. S. Department of Commerce, in providing the neutron research facilities used in this work.

\end{document}